\documentclass{emulateapj}

\usepackage{multirow}

\begin{document}

\title{Luminous Infrared Galaxies with the Submillimeter Array:
II. Comparing the CO~(3--2) Sizes and Luminosities of Local and High-Redshift 
Luminous Infrared Galaxies} 
\author {Daisuke Iono\altaffilmark{1,2}, 
Christine D. Wilson\altaffilmark{3},
Min S. Yun\altaffilmark{4}, 
Andrew J. Baker\altaffilmark{5}, 
Glen R. Petitpas\altaffilmark{6}, 
Alison B. Peck\altaffilmark{7}, 
Melanie Krips\altaffilmark{6},
T. J. Cox\altaffilmark{6}, 
Satoki Matsushita\altaffilmark{8}, 
J. Christopher Mihos\altaffilmark{9}, 
Ylva Pihlstrom\altaffilmark{10} 
} 
\altaffiltext{1}{Institute of Astronomy, The University of Tokyo, 2-21-1 Osawa, Mitaka, Tokyo 181-0015, Japan}
\altaffiltext{2}{Current address; National Radio Observatory, NAOJ, Minamimaki, Minamisaku, Nagano, 384-1305, Japan; d.iono@nao.ac.jp}
\altaffiltext{3}{Department of Physics and Astronomy, McMaster University, Hamilton, ON L8S 4M1, Canada}
\altaffiltext{4}{Department of Astronomy, University of Massachusetts, Amherst, MA 01003}
\altaffiltext{5}{Department of Physics and Astronomy, Rutgers, the State University of New Jersey, 136 Frelinghuysen Road, Piscataway, NJ 08854-8019}
\altaffiltext{6}{Harvard-Smithsonian Center for Astrophysics, Cambridge, MA 02138}
\altaffiltext{7}{Joint ALMA Office, Avda El Golf 40, piso 18, Santiago 7550108 Chile}
\altaffiltext{8}{Academia Sinica Institute of Astronomy and Astrophysics, Taipei 106, Taiwan}
\altaffiltext{9}{Department of Astronomy, Case Western Reserve University, 10900 Euclid Avenue, Cleveland, OH 44106}
\altaffiltext{10}{Department of Physics and Astronomy, University of New Mexico, Albuquerque, NM 87131}

\begin{abstract}
  We present a detailed comparison of the CO~(3--2) emitting molecular gas 
  between a local sample of luminous infrared galaxies (U/LIRGs) and a high 
  redshift sample that comprises submm selected galaxies (SMGs), 
  quasars, and Lyman Break Galaxies (LBGs).  
  The U/LIRG sample consists  of our recent CO~(3--2) survey 
  using the Submillimeter Array while 
  the CO~(3--2) data for the high redshift population are obtained from the
  literature.  We find that the  L$^{'}_{\rm CO(3-2)}$ and  
  L$_{\rm FIR}$ relation is correlated over five orders of magnitude, 
  which suggests that the molecular gas traced in CO~(3--2) emission is a 
  robust tracer of dusty star formation activity.  The near unity slope 
  of $0.93 \pm 0.03$ obtained from a fit to 
  this relation suggests that the star formation efficiency is constant   
  to within a factor of two  
  across different types of galaxies residing in vastly different epochs.    
  The CO~(3--2) size measurements suggest that  
  the molecular gas disks in local U/LIRGs (0.3 -- 3.1~kpc)
  are much more compact than the SMGs (3 -- 16~kpc), and that the
  size scales of SMGs are comparable to the 
  nuclear separation (5 -- 40~kpc) of the widely separated nuclei 
  of U/LIRGs in our sample.  
  We argue from these results that the SMGs studied here 
  are predominantly intermediate 
  stage mergers, 
  and that the wider line-widths arise from the violent merger of two
  massive gas-rich galaxies taking 
  place deep in a massive halo potential.  
%
\end{abstract}

\keywords{galaxies: formation, galaxies: starburst, cosmology: observations, galaxies: high redshift, submillimeter }

\section{Introduction}

Ultra/Luminous Infrared Galaxies (U/LIRGs) are sources 
that emit large amounts of flux in the  far-infrared (FIR) bands,
with FIR luminosities 
in the range 10$^{11-12}$~L$_{\odot}$ (for LIRGs) or even larger 
($>10^{12}$~L$_{\odot}$ for ULIRGs) 
\citep[see reviews by][]{sm96,lonsdale06}.  These galaxies
were discovered by the \textit{Infrared Astronomical Satellite} (IRAS) 
to be fairly rare in the local universe, but show
significant increase as a function of redshift \citep{hacking87}.
Followup optical studies of nearby U/LIRGs have revealed that the morphologies
of a significant fraction of them resemble interacting/merging systems 
\citep{armus87}, mostly powered by starbursts with increasing contribution 
from Active Galactic Nuclei (AGN) for IR-brighter galaxies \citep{kim98}.
These galaxies harbor large amounts of molecular gas \citep{sanders86} that is
often concentrated near the compact nuclear regions 
\citep{downessolomon98,bryant99}. These observational findings were 
analyzed in conjunction with numerical simulations that date back as far  
as the 1970's \citep{toomre72}, further advanced with the inclusion of 
gas dynamics and star formation recipes in the 1990's 
\citep{barnes96,mihos96} and with realistic feedback mechanisms implemented in 
the 2000's \citep{springel01,springel05,cox06}.  As a result, 
it is widely believed that 
tidally triggered gas compression and subsequent starbursts 
(and/or AGN fueling) heat the surrounding dust, are reprocessed
into far-infrared emission, and result in the intriguing U/LIRG phenomenon
we observe in the local as well as the high redshift universe.

Since the late 1990's, with the advent of sensitive mm/submm bolometer cameras 
mounted on single dish telescopes, observations 
of the blank sky have revealed the ubiquitous presence of  
unresolved sources that appear to 
account for a substantial fraction of the total infrared background 
radiation \citep[see][for a review]{lagache05}. 
By using the correlation between radio and FIR \citep{condon92}, 
interferometric radio observations toward 
these sources and subsequent followup optical spectroscopy 
have provided substantial evidence that the origin of this 
emission is dusty star-forming galaxies that reside 
at $z \geq 2$ \citep{chapman04b}.  
The FIR luminosities and
molecular gas properties in these submillimeter selected  
galaxies (SMGs) are generally an order of magnitude 
larger than the local U/LIRGs \citep{greve05,tacconi06,tacconi08}.
While the FIR luminosities in local U/LIRGs appear to be powered primarily 
by starbursts and in some cases a central AGN \citep[e.g.][]{genzel98},  
recent studies have suggested that 
the AGN contribution to the FIR luminosity is  
negligible in the high redshift SMGs \citep{pope06, pope08}. 

%

It has long been proposed that U/LIRGs and quasars are linked through 
dynamical evolution of the host systems, likely by a merger event 
\citep{sanders88, hopkins05}.  The initial
stages of the merger could be dominated by dusty starbursts 
(i.e. U/LIRGs and SMGs), until the central nuclear activity is stimulated by 
massive gas accretion.
When the cold gas is consumed either by starbursts or accretion onto the
central AGN, the quasar may become visible as 
intervening dusty high column density
material along our line of sight clears away.  
This merger evolution scenario is partially supported 
by recent deep and high-resolution imaging of quasar host 
galaxies \citep{dunlop03,sanchez04,veilleux06,zakamska06}, where morphological 
signatures of interactions/mergers and spheroidal systems are evident.

Since the important fuel of the activity, whether it be starbursts or AGN,
is warm and dense gas, a large sample of molecular line data in 
IR bright galaxies and quasars at different epochs  
is needed.  This data set allows us to test whether the evolutionary 
scenario is consistent with observations in the 
local universe, and then to test whether it holds true at high redshifts. 
Single dish and interferometric 
observations of the CO~(1--0) emission in local U/LIRGs have been carried
out previously, providing significant evidence that the amount of gas in these
systems is massive, and highly concentrated in the 
nucleus \citep[e.g.][]{downessolomon98}.  Similar observations have been 
conducted in CO emission of the redshifted high-$J$ transition lines from 
SMGs/quasars \citep[see][for a review]{svb05}, 
but these observations are difficult due to the requirement of 
a precise knowledge of the redshift and the lack of sensitivity at mm/submm
bands, and therefore the sample is biased toward the brightest (or
gravitationally lensed) sources.

The majority of the molecular gas detections in the bright SMGs/quasars 
thus far
from $z \gtrsim 2$ sources  
are CO~(3--2) (or higher J) line redshifted to the mm bands. 
However, directly comparing 
the properties of the CO~(3--2) emission in the SMGs/quasars 
with the CO~(1--0) emission in the local galaxies can be significantly
biased by excitation 
effects\footnote{T$_{ex} = 33$~K and n$_{crit} = 3.6 \times 10^{4}$~cm$^{-3}$ 
for CO~(3--2) whereas T$_{ex} = 6$~K 
and n$_{crit}= 2.2 \times 10^{3}$~cm$^{-3}$ for CO~(1--0)}.  
Thus, to properly assess the characteristics of gas in 
U/LIRGs and high redshift SMGs/quasars, 
a thorough comparison of a single $J$ transition emission line is needed. 
We have carried out a program at the Submillimeter Array 
(SMA)\footnote{The Submillimeter 
Array is a joint project between the Smithsonian Astrophysical Observatory 
and the Academia Sinica Institute of Astronomy and Astrophysics, and is 
funded by the Smithsonian Institution and the Academia Sinica.}\citep{ho04} to 
observe 14 U/LIRGs in the local universe in the CO~(3--2) and CO~(2--1) lines 
\citep[][Paper~I hereafter]{wilson08}.  
Because the critical density of CO~(3--2) emission is relatively high,
it is considered as an important tracer of the physical condition of the  
star forming gas and the associated kinematics.  
This common belief, however, 
has not been fully tested observationally because of the lack of high 
resolution observations of nearby and distant sources.  
Recent high resolution studies of local U/LIRGs show,  
in some cases, that the distribution and the peak of the CO~(3--2) emission 
could be different from the lower transition CO \citep[e.g.][]{iono04}.

In order to conduct an analysis that is not biased by excitation effects,
we present a detailed comparison of the CO~(3--2) emitting molecular gas 
between a local sample of U/LIRGs and a high 
redshift sample  that comprises SMGs, quasars, and  
two Lyman Break Galaxies (LBGs) that have been recently detected in CO~(3--2).
We also include CO~(3--2) data from a sample of local quiescent galaxies 
in some of the analyses.  
The U/LIRG sample consists  of our recent CO~(3--2) survey 
using the SMA (Paper~I) while the CO~(3--2) data for the high redshift 
and local quiescent galaxies are obtained from the literature.

In \S2, we present the sample data and the formalism for deriving the
relevant parameters.  In \S3, we investigate the correlation between CO~(3--2) 
and FIR luminosities in a variety of sources, and offer possible 
interpretations of the derived slope and the scatter 
seen in the correlation. We compare  
the CO~(3--2) source size in different sources and 
offer
clues to the merger stage of SMGs and quasars in \S4.  
We present our conclusions in \S5.
We adopt H$_0$ = 70~km~s$^{-1}$~Mpc$^{-1}$, 
$\Omega_M$ = 0.3, $\Omega_{\Lambda}$ = 0.7 for all of the 
analysis throughout this paper.

\section{Sample Data and Derived Quantities}

Observational details and the properties of the local U/LIRG data are
described  in Paper~I.  
The U/LIRG sample consists of 14 sources, of which six are 
interacting/merging galaxy pairs whose nuclei are resolved into 
multiple components (``pre-coalescence'' mergers), and eight are 
mergers whose 
nuclei are unresolved even with space-based optical images 
(``post-coalescence'' mergers).  
For the remainder of the paper, the local infrared bright galaxies   
with L$_{\rm FIR} > 10^{12}$~L$_{\odot}$ are referred to as 
ULIRGs, whereas galaxies with 
L$_{\rm FIR} < 10^{12}$~L$_{\odot}$ are referred to as LIRGs.  
The term U/LIRG is used when referring to the entire sample 
presented in Paper~I, regardless of the FIR luminosity.

Interferometric CO~(3--2) data toward SMGs were obtained from 
published results in \citet{neri03}, \citet{sheth04},
\citet{greve05}, \citet{tacconi06,tacconi08} and \citet{knudsen08}.
A few sources that were observed in \citet{greve05} were also observed
by \citet{tacconi06} at higher angular resolution, and we adopt the  
latter for our analysis.  
In addition, CO~(3--2) emission in MIPS-J1428 \citep{iono06a}, 
a submm bright  HyLIRG (Hyper Luminous InfraRed Galaxy; 
L$_{\rm FIR} > 10^{13}$~L$_{\odot}$) at $z=1.3$, 
was also included in our SMG sample. 
The CO~(3--2) data for the quasars were compiled from \citet{hainline04}, 
\citet{walter04}, \citet{svb05}, and \citet{coppin08}.
The two  CO~(3--2) detected LBGs, MS~1512-cB58 (cB58 hereafter) 
and J213512-010143 (a.k.a. the ``cosmic eye'', J213512 hereafter), 
were obtained from \citet{baker04} and \citet{coppin07}, respectively

The far infrared luminosities (L$_{\rm FIR}$) are obtained from the IRAS 
Revised Bright Galaxy Sample (BGS) \citep{sanders03} for all of the U/LIRGs.  
The  L$_{\rm FIR}$ of SMGs/quasars are derived by 
fitting a theoretical starburst spectral energy distribution (SED)
\citep{efstathiou00}  
to the observed FIR to radio flux densities \citep{yun02}.  The 
same SEDs were used to compute the L$_{\rm FIR}$  of the three local
ULIRGs in our sample, and we found that 
they are consistent with the \citet{sanders03} values to within $<50$\%.
The L$_{\rm FIR}$ of the LBGs are obtained from \citet{baker04} 
and \citet{coppin07}.
The sample sources along with various physical properties  
used in this study are presented in Table~\ref{table1}.

We derive the following physical properties for the analysis of this paper.
The derived physical properties of all of the sources are presented 
in Table~\ref{table2}.

(1) The CO~(3--2) luminosity, L$^{'}_{\rm CO(3-2)}$ [K~km~s$^{-1}$~pc$^2$], 
is derived using
\begin{equation}
L^{'}_{\rm CO(3-2)} = 3.25 \times 10^{7}~S_{\rm CO(3-2)}~\nu^{-2}_{obs}~(1+z)^{-3}~D^2_L. 
\end{equation}
where $S_{\rm CO(3-2)}$ is the integrated CO(3--2) intensity in 
Jy~km~s$^{-1}$, $\nu_{obs}$ is the observed frequency in GHz, $D_L$ is the
luminosity distance in Mpc, and $z$ is the redshift of the 
source \citep{svb05}.  

(2) The full width at half maximum (FWHM) of the CO~(3--2) line are estimated 
using the integrated spectra shown in Fig.~25--29 of Paper~I, and obtained 
from the published results for the SMGs, quasars, and LBGs.
 


(3) The source sizes 
of U/LIRGs are the deconvolved FWHM
diameters   
derived by fitting a 2 dimensional Gaussian to 
the CO~(3--2) integrated intensity maps.  
When the deconvolution failed in some of the sources (mainly due to 
low S/N),  the major and minor FWHM axes 
obtained from the Gaussian fits in the image domain are given as upper
limits to the source size.
All of the U/LIRGs are
resolved with the SMA beam, except for IRAS~17208-0014  
whose deconvolution produces a point source.  However, visual inspection
of its integrated intensity map in Figure~2 of Paper~I suggests that 
the CO~(3--2) emission appears to have a bright compact component, as well 
as a resolved extended structure.   

The published sizes are used for the 6 SMGs that are resolved by the
interferometer.  Most of the quasars are unresolved 
except for SDSS~J1148+5251, IRAS~F10214+4724, and Cloverleaf.  High
angular resolution VLA  
observations \citep{walter04} have successfully resolved the 
structure in the $z=6.4$ quasar SDSS~J1148+5251, whereas OVRO observations 
\citep{yun97} in conjunction with a lensing model have
inferred a source size of $0\farcs3$ for the highly lensed $z=2.6$ quasar
Cloverleaf.
For the remaining unresolved SMGs, quasars, and LBGs, the minor axis
of the beam (FWHM) is used as an upper limit to the source size.


Because the physical resolution  
    of the high redshift population is worse than the local LIRGs, some of 
    the widely separated sources that are resolved with the SMA may not 
    be resolved using $0\farcs8$ resolution when they are placed at $z=2$.  
    We have tested this by convolving the CO~(3--2) images of all of
    the widely separated pairs 
    with $0\farcs8$ ($\sim 8$~kpc at $z=2$) resolution, 
    and found that all the galaxies are clearly resolved into multiple
components except for 
Arp~299, which is only marginally resolved. 

(4) The star formation efficiency (SFE),  
L$_{\rm FIR}$/L$^{'}_{\rm CO(3-2)}$, is derived
by taking the ratio between FIR luminosity and CO~(3--2) luminosity.  The
ratio between FIR luminosity and molecular gas mass 
(i.e. L$_{\rm FIR}$/M$_{\rm H_2}$) is often used to infer the SFE of a 
galaxy, but we opted to use the luminosity ratio 
(L$_{\rm FIR}$/L$^{'}_{\rm CO(3-2)}$) in this study 
to avoid introducing additional ambiguities through  variation and 
uncertainties pertaining to the 
conversion from L$^{'}_{\rm CO(3-2)}$ to M$_{\rm H_2}$.


We caution that 
the CO~(3--2) properties at 
low and high redshifts are generally lower limits for the following 
reasons. 
In synthesis imaging, incomplete uv coverage results in decreasing sensitivity 
to extended structure. 
While the SMA observations are sensitive to structure  
with size scales of $\lesssim 12'' - 16''$ 
(7 -- 10~kpc on average for these galaxies), 
the missing flux calculated by comparing to single dish 
data is 2 -- 66\% (Paper~I).  
Thus the diffuse and extended CO~(3--2) emission
would be undetected in the SMA observations, 
and the values derived in this paper are lower 
limits to the physical quantities intrinsic to the source.  
On the other hand, observations of high redshift galaxies are limited in 
both sensitivity and angular resolution.  The relatively 
coarse angular (physical) resolution of $\lesssim 0\farcs8$ ($\sim 6$~kpc) 
allows us to analyze images that do not suffer significantly from the 
missing-flux problem.  However, the high redshift population 
suffers from limited sensitivity (with mass sensitivity of 
$\sim 10^{10}$~M$_{\odot}$), and any extended CO~(3--2)  emission
could be buried in the noise.  
Therefore, the physical quantities derived here from CO~(3--2) data
for both the low and high 
redshift populations are likely lower limits.

\section{The CO~(3--2) Luminosity and the Star Formation Efficiency}

The luminosity of the CO~(3--2) emission line can be a
direct measure of the amount of dense molecular gas  
fueling nuclear (or extended) star formation and/or the central AGN.  
The average CO~(3--2) luminosities (L$^{'}_{\rm CO(3-2)}$) derived 
in U/LIRGs, SMGs, and quasars are $(2.6 \pm 0.5) \times 10^{9}$,
$(4.4 \pm 1.1) \times 10^{10}$ and   
$(5.0 \pm 1.0) \times 10^{10}$~K~km~s$^{-1}$~pc$^{2}$, respectively.
The L$^{'}_{\rm CO(3-2)}$ for the two LBGs are 
$4.4 \times 10^{8}$ and   
$2.9 \times 10^{9}$~K~km~s$^{-1}$~pc$^{2}$ for cB58 and J213512, respectively.
The average CO~(3--2) luminosity in the U/LIRGs is more than an order of 
magnitude lower than in the SMGs and quasars, but comparable to the two LBGs.
Under the assumption that the powering source of the large L$_{\rm FIR}$ 
is mostly from starbursts (see \S3.1), the comparison between
L$^{'}_{\rm CO(3-2)}$  
and  L$_{\rm FIR}$ relates the amount 
of available dense molecular gas to the amount of 
current massive star formation traced in FIR dust emission.
This analysis assumes an environment where 
the gas is sufficiently dense and warm to thermally 
(or radiatively) populate the J=3 rotational energy level of carbon monoxide,
and that the gas and dust are coupled spatially.  
We investigate the relation between L$^{'}_{\rm CO(3-2)}$ and 
L$_{\rm FIR}$ in the following subsection.

\subsection{The L$^{'}_{\rm CO(3-2)}$ -- L$_{\rm FIR}$ Relation}

\begin{figure}
  \plottwo{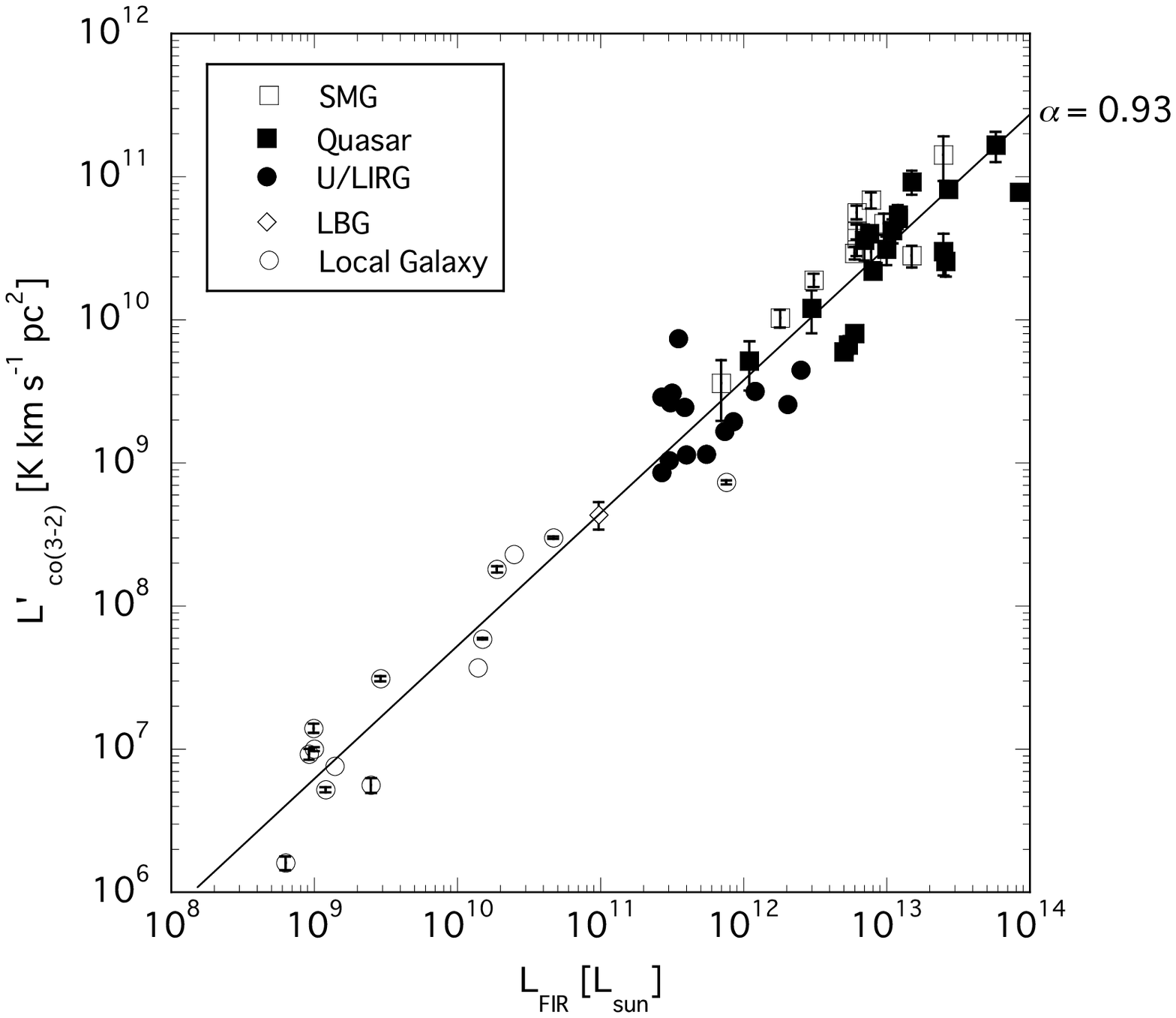}{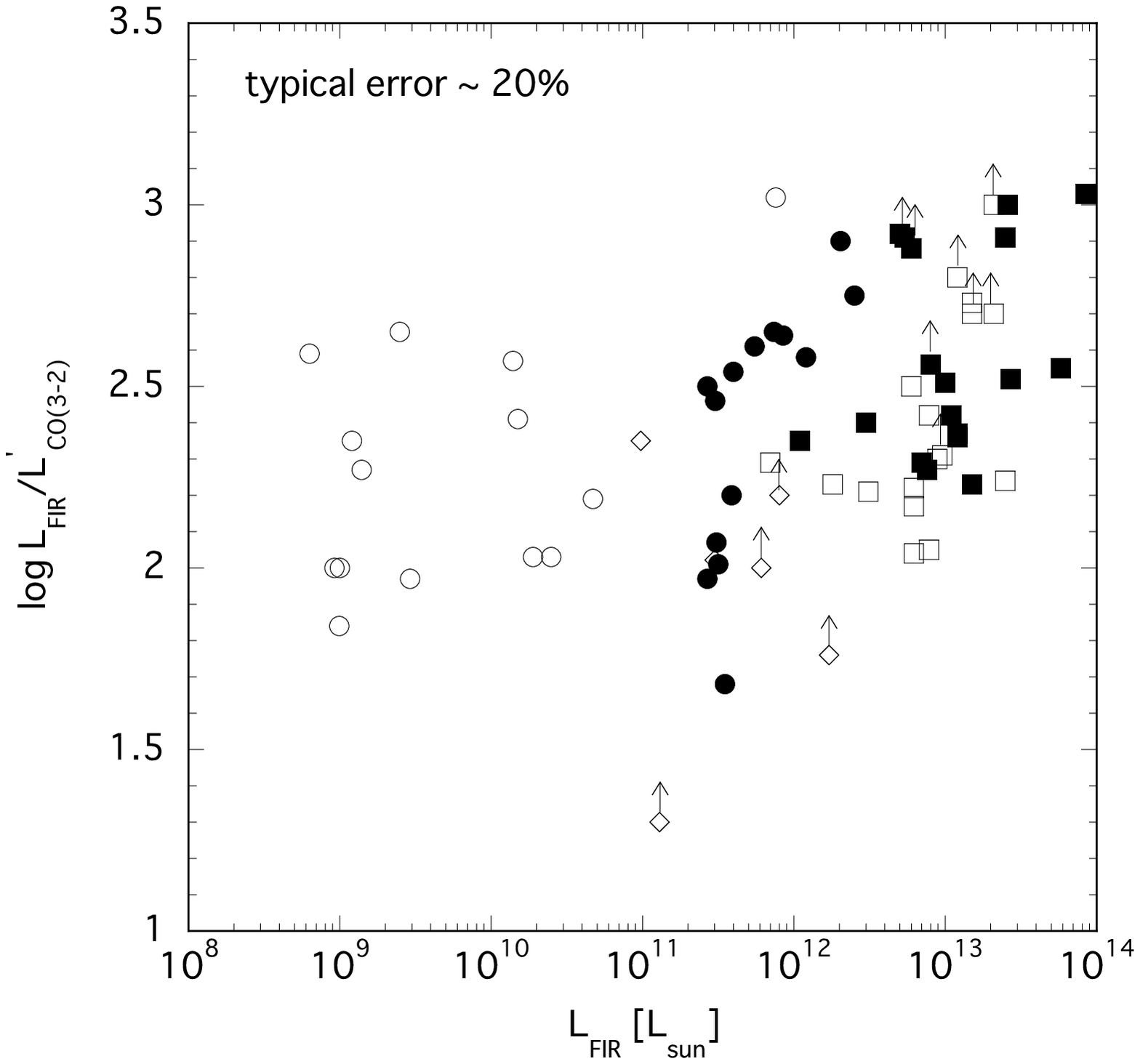}
  \caption{
    ($left$) The FIR Luminosity vs. L$^{'}_{\rm CO(3-2)}$ for all four 
    populations.  We have additionally plotted the same relation for the 
    local galaxies. The fit to all of the points is shown in the solid line.  
    ($right$) The L$_{\rm FIR}$ to L$^{'}_{\rm CO(3-2)}$ ratio vs. the 
    FIR luminosity plotted in logarithmic scales.  
    The symbols are the same as in $left$.  We have additionally 
    plotted the same relation for sources that were  
    undetected in CO~(3--2) using data published in \citet{greve05}, 
    \citet{tacconi08}, \citet{coppin08} and \citet{hatsukade09}.  
    The FIR luminosities of the optically selected galaxies (classified as 
    LBGs in this plot) in 
    \citet{tacconi08} were estimated by taking the SFRs from \citet{genzel06}
    and \citet{erb06}, and using the conversion given in \citet{kennicutt98}.
  }
  \label{fig1}
\end{figure}

In Figure~\ref{fig1}~($left$), we compare the CO~(3--2) 
line luminosity with the FIR luminosity for all four sample populations.
In addition, we have plotted the same luminosity relation for a sample of 
local galaxies published in \citet{mauersberger99} and \citet{komugi07} to
increase the luminosity range.  
The L$^{'}_{\rm CO(3-2)}$ -- L$_{\rm FIR}$ relation including all four 
populations (and local galaxies) 
is correlated over five orders of magnitude in 
luminosity, which suggests that the molecular gas seen in CO~(3--2) 
emission is a robust tracer of  star formation activity.  
A least-squares fit of the form 
log~L$^{'}_{\rm CO(3-2)} =  \alpha$ log~L$_{\rm FIR} + \beta$
gives $\alpha = (0.93 \pm 0.03)$ and $\beta = (-1.50 \pm 0.33)$.
Excluding the local galaxies or the quasars, 
which may have significant dust heating due to the central AGN, 
does not change the slope with significance. 
One potential concern here is that the large range in distance may artificially
produce a correlation in log-log space. To check whether the
correlation seen here is statistically robust, we employed  
the partial Kendall $\tau$-coefficient with censored data \citep{akritas96} 
where we used the luminosity distance as a test variable.  
This analysis yields a probability of $7.3\times 10^{-5}$ that these
two variables  
will produce a false correlation after the effect of distance is removed, 
providing solid statistical evidence of correlation in logarithmic space.  

The near unity slope derived from the L$_{\rm FIR}$ -- L$^{'}_{\rm CO(3-2)}$ 
relation suggests that the efficiency of converting CO~(3--2) emitting 
molecular gas to massive stars (i.e. SFE) is, 
within a factor of two (i.e. from the standard deviation; see below), 
nearly uniform across different types of galaxies residing in vastly 
different epochs.  To illustrate this better, we plot  
the SFEs against the FIR luminosities
in Figure~\ref{fig1}~($right$).
The average SFEs in the various galaxy samples range from a low of 170
for the SMGs and LBGs to 250 for the LIRGs to 430 for the quasars and
a high of 580 for the ULIRGs.
The combined average of the SMGs, quasars, LBGs and U/LIRGs 
is $375 \pm 42$~(standard deviation = 327)
~L$_\odot$ (K~km~s$^{-1}$~pc$^2$)$^{-1}$. The average SFE of the SMGs is
lower than the U/LIRGs by a factor of two, 
while a slightly higher average SFE seen in the quasars (and also the
ULIRGs) is possibly attributable to
higher dust temperature as a result of dust heating by the central AGN.   
Interestingly, 
the SFEs of the SMGs are more consistent with the widely 
separated LIRGs (and the LBGs and local galaxies) than the ULIRGs and quasars. 
We will return to this point in our discussion of the evolutionary
status of the SMGs in \S4.2.
The SFE and L$_{\rm FIR}$  properties of the LBGs are also
generally similar to the 
LIRGs, but this sample is currently limited 
to only two galaxies.

In U/LIRGs, the exact fraction of the AGN/starburst  
contribution to the FIR luminosity appears to depend on the source
\citep{armus07}.   Half of the fourteen U/LIRGs in our local
sample show evidence for an AGN, but only in the two ULIRGs Mrk231 and
Mrk273 is it possible that the AGN makes a significant contribution to
the bolometric luminosity (Paper~I, and references therein).
This variation suggests that the scatter in the SFEs  
of U/LIRGs not only reflects the physical characteristics of each galaxy, 
but also the AGN heating of L$_{\rm FIR}$ in some sources.
However, we mention the overall trend that  
the FIR output in local IRAS sources 
is dominated by starbursts and not AGN \citep{yun01}, as is evident   
from the robustness of the empirical radio-FIR correlation.
In addition, while there is evidence from X-ray surveys 
that a good fraction of SMGs contain AGNs \citep[$\sim 80\%$;][]{alexander05}, 
the energetic contribution of the AGN to the FIR output in most of the  
SMGs appears to be minor \citep{pope08,valiante07}. Indeed, there is
even evidence that most of the far-infrared luminosity in quasars
comes from a central starburst \citep{lutz07,wang07,coppin08},
although there are exceptions \citep{weiss07a}.

In Paper~I, it was found that the relation between 
M$_{\rm H_2}$~(or L$^{'}_{\rm CO(3-2)}$) 
and L$_{\rm FIR}$ was not correlated with any significance, leading to
an argument that the degree of gas concentration
determines the level of star formation activity rather than the 
overall amount of molecular gas mass available in the galaxy.
Due to large scatter in the SFEs of LIRGs
(see Figure~\ref{fig1}~($right$)), 
the correlation indeed breaks down if we only consider the U/LIRGs in the fit. 
The LIRGs that have particularly low SFEs ($\lesssim 150$) 
are NGC~6240, Arp~193, VV~114, NGC~5257/8, and NGC~5331.
Optically thin CO~(3--2) is suggested from previous analysis
on NGC~6240 \citep{iono07}, and therefore the CO~(3--2) emission 
may be over-luminous with respect to other LIRGs with 
optically thick CO~(3--2) emission.  
Although Arp~193 is apparently a late stage merger, VV~114, NGC~5257/8,
and NGC~5331 show large nuclear separation suggestive of early 
stages of interaction.  The SFEs of these galaxies are 
93 -- 117~L$_{\odot}~$(K~km~s$^{-1}$~pc$^2$)$^{-1}$, 
which is a factor of three lower than the average of the entire sample.  
The factor of two lower average SFE seen in 
LIRGs (which have a mix of compact and widely separated sources) compared to 
ULIRGs (which are mainly compact) may suggest that the global SFE is lower in 
early to intermediate stage pre-coalescence mergers than in the centers of 
more advanced stage mergers, 
likely because the gas in intermediate stage mergers 
has not settled in a steady state in the rapidly evolving host stellar 
potential. 

Finally, we note that the average SFE in local U/LIRGs is a factor of 
seven larger than L$_{\rm FIR}$/L$^{'}_{\rm CO(2-1)} \sim 50$ 
found in two $z \sim 1.5$ BzK-color selected ULIRGs \citep{daddi08}. 
However, L$_{\rm FIR}$/L$^{'}_{\rm CO(2-1)}$ 
may be significantly 
lower than  L$_{\rm FIR}$/L$^{'}_{\rm CO(3-2)}$ in the same galaxy 
owing to excitation effects, and future 
CO~(3--2) observations of the same $z \sim 1.5$ BzK selected ULIRGs are  
necessary to compare the true CO~(3--2) properties of molecular gas in 
ULIRGs at this redshift. \citet{tacconi08} have obtained upper limits
to the CO~(3--2) emission in three optically selected star forming
galaxies similar to the BzK sample; however, the average lower limit
to their SFE \citep[$>90$, where we have estimated L$_{\rm FIR}$ from their
measured star formation rates using the relation in][]{kennicutt98} implies 
that the local U/LIRGs have at most a factor of three times larger
SFE. Upper limits to the CO~(3--2) emission in 5 SMGs and 3 quasars
\citep{greve05,coppin08} give lower limits to their SFEs of $> 160$ to
$>1000$. While most of these limits lie above the average SFE shown in
Figure~\ref{fig1}~($right$), 
significantly more sensitive CO~(3--2)
observations of these sources would be needed to show conclusively 
that they deviate
significantly in their properties from the detected SMGs and quasars.

\subsection{Slope in the  L$_{\rm FIR}$ -- L$^{'}_{\rm CO}$ Relation}

The CO~(3--2) derived slope ($\alpha_{\rm CO(3-2)} = 0.93 \pm 0.03$) in the  
L$_{\rm FIR}$ -- L$^{'}_{\rm CO}$ relation is significantly steeper 
than a previous 
compilation which includes different excitation lines from local to 
high redshift sources 
\citep[i.e. $\alpha_{\rm CO} = 0.62 \pm 0.08$;][]{greve05}.  
This is expected because the molecular ISM 
in most of the SMGs is subthermally excited
in CO transitions higher than J = 4--3 
 \citep{weiss07}, and 
their sample of local U/LIRGs  mainly consists of observations of J = 1--0.
A study that investigates only the CO~(3--2) line 
in local galaxies with FIR luminosities in the range 
$10^{9}$--$10^{12}$~L$_{\odot}$ was conducted by \citet{yao03}, 
where they find $\alpha_{\rm CO(3-2)} = 0.70 \pm 0.04$.
A more recent study \citep{narayanan05} that relates the 
CO~(3--2) to IR luminosity in a similar luminosity range 
suggests a near linear correlation ($\alpha_{\rm CO(3-2)} = 1.08 \pm 0.07$).
While there is variation in the slope among different studies, 
the slope between the CO~(1--0) line and the FIR luminosity 
is even flatter \citep[$\alpha_{\rm CO(1-0)} = 0.58 \pm 0.07$;][]{yao03} than 
that derived using the CO~(3--2) emission alone.
The relatively flat CO~(1--0) slope suggests that  
the CO~(1--0) derived SFE increases with increasing 
FIR luminosity \citep[i.e.;][]{sanders88}, 
and the molecular gas traced in
CO~(1--0) emission is not a linear tracer of star formation activity.  

It has been demonstrated that the HCN emission, 
which is a higher density gas tracer than the commonly used CO~(1--0) 
emission, produces a linear correlation with FIR luminosity 
\citep[$\alpha_{\rm HCN(1-0)} = 1.00\pm0.05$;][]{gao04}. 
Similar recent studies \citep{gao07,riechers07,carpio08} 
that include HCN observations from high redshift galaxies have shown, 
however, that L$_{\rm FIR}$/L$^{'}_{\rm HCN}$ 
in the most distant galaxies is larger 
than the local sample by at least a factor of two, 
and the resultant fit between L$^{'}_{\rm HCN(1-0)}$ and L$_{\rm FIR}$ 
becomes flatter \citep[$\alpha_{\rm HCN(1-0)} = 0.81\pm0.06$;][]{carpio08} 
when high redshift galaxies are included.
These new HCN results suggest that
the dense gas fraction in bright high redshift galaxies is even higher than
in the local IR bright galaxies.

\begin{figure}
  \plotone{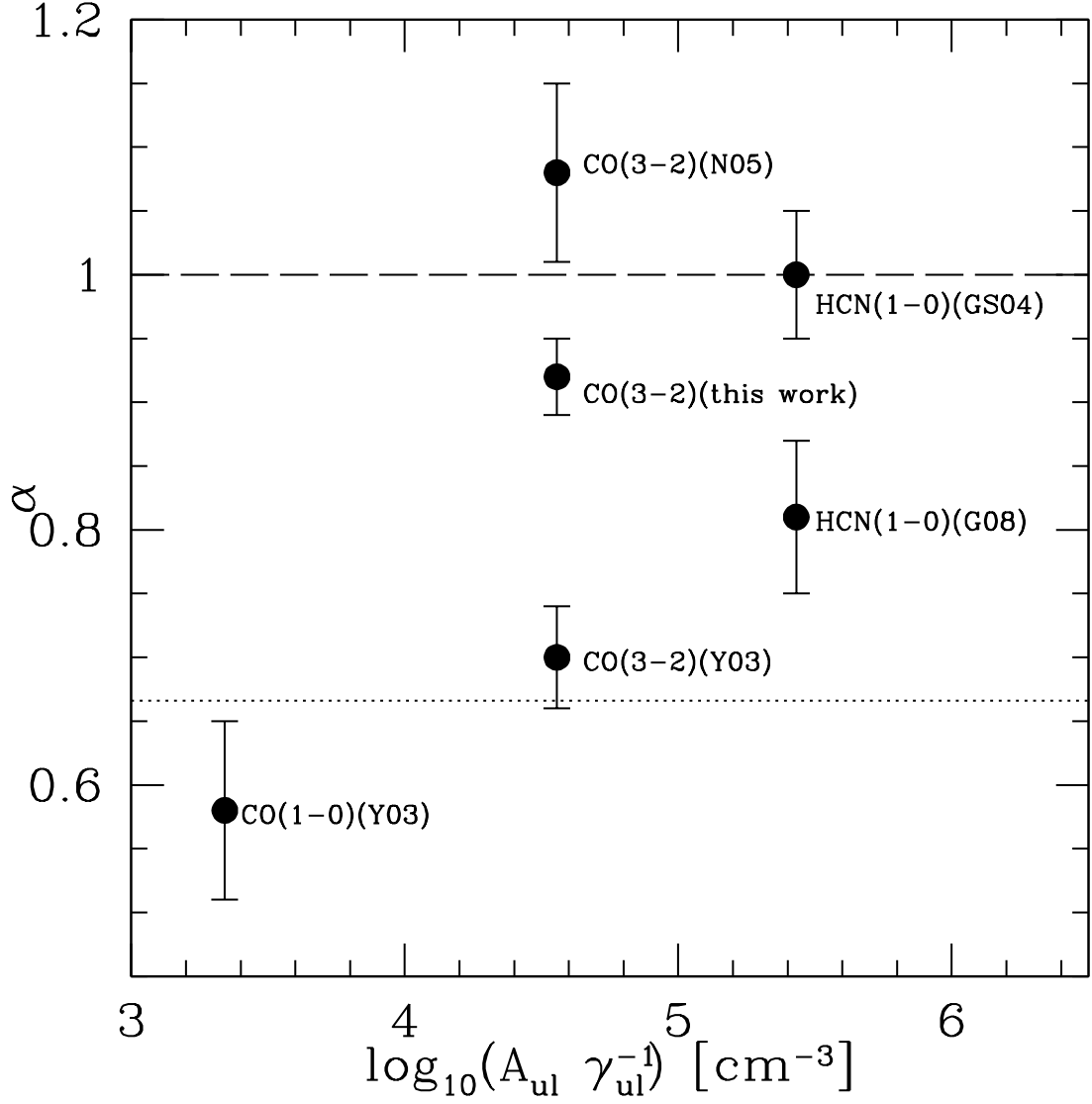}
  \caption{
    The index $\alpha_{mol}$ versus the critical density 
    ($A_{ul}$$\gamma^{-1}_{ul}$) 
    in the CO~(1--0), CO~(3--2), and the HCN~(1--0) lines, where 
    $A_{ul}$ is the Einstein A coefficient in units of s$^{-1}$ 
    and $\gamma^{-1}_{ul}$ is the 
    collision rate coefficient in units of cm$^{3}$~s$^{-1}$.  
    The collision rates ($\gamma_{ul}$) are obtained from \citet{schoier05}, 
    and the average of T = 10,20,30,50~K are used.  The dotted and short
    dashed lines represent the theoretical prediction for CO~(1--0) and 
    HCN~(1--0) respectively from \citet{krumholz07}.  References: 
    \citet{yao03}~(Y03), \citet{narayanan05}~(N05), \citet{gao04}~(GS04),
    \citet{carpio08}~(G08).
  }
  \label{fig2}
\end{figure}

In Figure~\ref{fig2}, we show the different slopes ($\alpha_{mol}$) 
plotted against the critical densities of the CO~(1--0), 
CO~(3--2) and HCN~(1--0) transitions.  We calculate the 
critical densities from $A_{ul} \gamma^{-1}_{ul}$ where 
$A_{ul}$ is the Einstein $A$ coefficient for the upper ($u$) to 
the lower ($l$) energy state and $\gamma_{ul}$ is the corresponding 
collision rate obtained from the compilation of \citet{schoier05}. 
A general trend that $\alpha_{mol}$ approaches unity for 
higher critical density tracers is evident from Figure~\ref{fig2}.

The slope derived from CO~(1--0) emission, $\alpha_{\rm CO(1-0)}$,  
could be $\sim 2/3$ if a fixed fraction of gas converts to stars 
each free-fall time \citep{kennicutt98}.
From simple theoretical considerations, \citet{krumholz07}
argue that, because the critical density of the CO~(1--0) line  
is lower than the higher density tracers such as HCN, 
the CO~(1--0) line traces the molecular gas that has densities 
close to the median density of a galaxy.  
On the other hand, molecules with high critical densities (such as 
the HCN~(1--0) line) trace the ISM with much higher densities. 
In this case, L$_{\rm FIR}$/L$^{'}_{\rm HCN(1-0)}$   
does not depend strongly on the mean density of the galaxy, 
and therefore the slope derived from the HCN~(1--0) measurements  
(i.e. $\alpha_{\rm HCN(1-0)}$) should be close to unity.  
They find, however, that the dependence 
of L$_{\rm FIR}$/L$^{'}_{\rm HCN(1-0)}$ to the 
mean density becomes more significant at higher ($10^{3-4}$~cm$^{-3}$) mean 
densities, suggesting that the slope can be flatter at the highest 
luminosity regime.   

We plot the two theoretical predictions 
for $\alpha_{\rm CO(1-0)}$~(dotted line) and 
$\alpha_{\rm HCN(1-0)}$~(short dashed line) 
in Figure~\ref{fig2} as horizontal lines. 
The observations for these two emission lines are consistent with the
theoretical predictions, and the new measurement by \citet{carpio08} 
is also consistent with the prediction by
\citet{krumholz07} that $\alpha_{\rm HCN}$ is less than unity 
when galaxies with higher mean densities 
(represented by high redshift sources) 
are included in the fit.  Our CO~(3--2) measurements, as well as those 
measured by others, fall between the CO~(1--0) and HCN~(1--0) derived 
slopes, but the work by \citet{krumholz07} was limited to an isothermal 
case, and therefore could not investigate higher transition lines 
that are excited at higher temperatures such as the CO~(3--2) line.

\subsection{Molecular Excitation and Large Velocity Gradient Modeling}

A theoretical investigation was performed by \citet{narayanan07b} 
to understand 
the dynamically evolving ISM of isolated and interacting galaxies.  
From their N-body/SPH simulations that include radiative transfer, 
they find that the total CO~(3--2) luminosity 
has a mix of contribution from the dense cores (sites of star formation) and 
the more extended molecular gas that is subthermally excited.
They argue 
that CO~(3--2) and FIR luminosities are linearly correlated under 
the assumption that the gas density and SFR density are related by 
a Schmidt law \citep{schmidt59} index of 1.5.   
Further, a notable scatter is seen  
in the low luminosity end of their CO~(3--2) -- FIR luminosity relation, 
which is caused by line trapping of the CO~(3--2) line 
by molecular gas surrounding the dense
star forming cores and suggests that subthermally excited CO~(3--2) gas can
provide a substantial contribution to the overall L$^{'}_{\rm CO(3-2)}$.

In order to investigate the CO~(3--2) excitation at different spatial
scales, we derived the total ($R_{total}$)
and peak ($R_{peak}$) CO~(3--2) to CO~(1--0) ratios in U/LIRGs using the 
CO~(1--0) data available in the literature. 
The results are presented in Table~\ref{table3}. The average $R_{total}$ 
is 0.48, ranging from 0.20 (Arp~55) to 1.25 (Arp~299W), whereas 
the average $R_{peak}$ is 0.96, ranging from 0.36 (NGC~5257) 
to 3.21 (NGC~6240).
On average, $R_{peak}$ is two times higher than $R_{total}$.
The two ratios are comparable for some of the 
compact sources such as IRAS~17208-0014, UGC~5101, Arp~299W, or NGC~5331N, 
but much larger in NGC~6240 (a factor of four larger).  
The lowest $R_{peak}$ is derived in the brightest ULIRG 
in our sample, IRAS~17208-0014.  The general inference of these results is 
that the CO~(3--2) transition is nearly 
thermalized (in the optically thick limit) 
in the inner 1 -- 3~kpc of the U/LIRGs, 
whereas the CO~(3--2) transition is subthermally populated in the 
extended outskirts. 
This result is consistent with the theoretical prediction by 
\citet{narayanan07b}, possibly suggesting that the scatter seen in the 
U/LIRGs in Figure~\ref{fig1} reflects the excitation 
characteristics of each galaxy.   Finally, we note that 
the $total$ CO~(3--2) emission in SMGs/quasars is nearly thermalized 
\citep{weiss07}, and this could be the reason for 
the tighter correlation seen in the SMGs in Figure~\ref{fig1}.

\begin{figure}
  \plotone{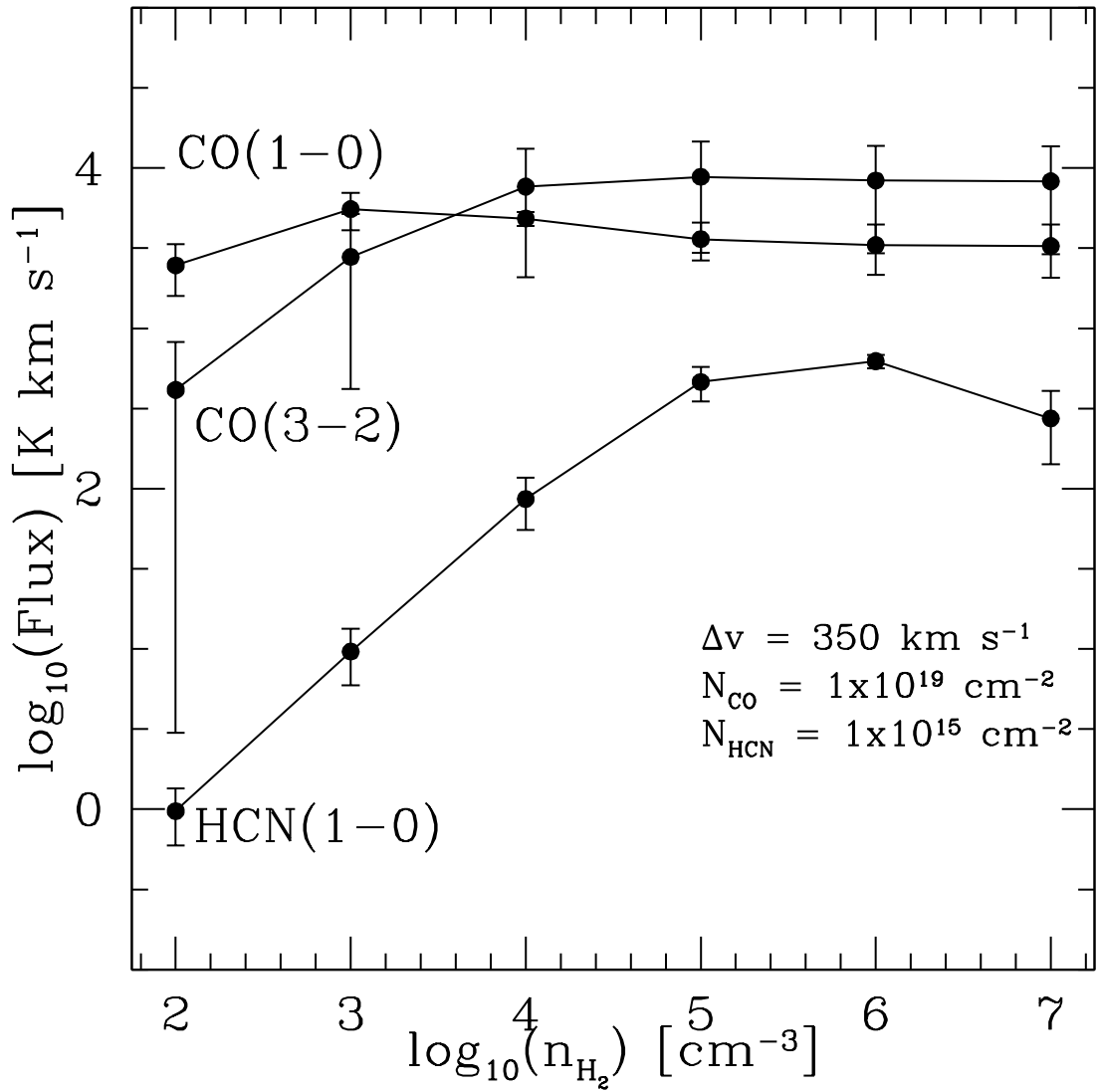}
  \caption{
    Large Velocity Gradient (LVG) modeling of the CO~(1--0), CO~(3--2) 
    and HCN~(1--0) lines.  The average of the 
    expected flux (in K~km~s$^{-1}$) is 
    shown as a function of H$_2$ density.  The error bars
    represent the range of possible fluxes when the kinetic temperature 
    is varied from 20, 40, 60, 80, and 100K.  The adopted LVG parameters 
    are shown in the lower right.  The column densities are motivated 
    by the observational evidence of N$_{\rm H_2} > 10^{23}$~cm$^{2}$ 
    in SMGs and U/LIRGs, and fractional abundance of CO and HCN to H$_2$ of 
    X$_{\rm CO} = 10^{-4}$ and X$_{\rm HCN} = 10^{-8}$.  We note that 
    changing N$_{\rm H_2}$ will shift the absolute flux values 
    for a given $n_{\rm H_2}$, but the general shape remains the same. 
    $\Delta V = 350$~km~s$^{-1}$ is the average CO~(3--2) FWHM in U/LIRGs.
  }
  \label{fig3}
\end{figure}

In order to obtain a better physical understanding of  
the observed variation in $\alpha_{mol}$ as well
as the observed scatter in the SFEs,  
we constructed a series of Large Velocity Gradient (LVG) models 
\citep{goldreich76} using the publicly available 
RADEX code \citep{vandertak07}.
Figure~\ref{fig3} shows the expected flux as a function of H$_2$ density
(see figure caption for the adopted input parameters).  Three key 
arguments can be made from this figure.  First, the rise in 
flux is nearly linear up to $\sim 10^4$~cm$^{-3}$ for CO~(3--2) and 
up to $\sim 10^5$~cm$^{-3}$ for HCN~(1--0), suggesting that these 
molecular transitions are good tracers of the density of the molecular 
medium up to these characteristic densities. The predicted flux beyond these 
characteristic values reaches an asymptotic value as the relevant lines
become thermalized (i.e. T$_{ex} \sim$~T$_{kinetic}$), thus predicting the 
gas densities from these lines alone becomes difficult in this regime. 
Second, the predicted 
flux of CO~(3--2) and CO~(1--0) lines cross at $\sim 10^{4}$~cm$^{-3}$ 
(i.e. flux ratio of unity), and the line ratio is less than unity 
for densities below this critical value.  
Lastly, the temperature dependence
of CO~(3--2) is significant compared to the CO~(1--0) 
line because the excitation temperature is higher in CO~(3--2).
At low densities ($n_{\rm H_2} \sim 10^2$~cm$^{-3}$), 
the observed CO~(3--2) flux can vary by two orders of magnitude if 
the temperature of the ISM varies from galaxy to galaxy.
Thus, at densities less than the critical density of the transition, 
collisional and radiative excitation both play key roles in populating 
the energy levels. 

Under the assumptions adopted for the LVG model used here, 
the observed variation in $\alpha_{mol}$ can be explained in the following way.
The low $\alpha_{\rm CO(1-0)}$ likely reflects the nature of CO~(1--0) 
emission as a low density tracer, and thus the higher density gas in U/LIRGs
do not further increase the observed CO~(1--0) flux, leading to the flattening
of $\alpha_{\rm CO(1-0)}$.  The CO~(3--2) emission is likely a tracer of 
higher density gas, but its flux is also sensitive to temperature, most notably
in the low density end.   
Therefore, the relatively large scatter in the U/LIRGs and local galaxies in  
Figure~\ref{fig1}~($left$) 
could be explained by intrinsic variation in the average gas kinetic 
temperature and density, 
reflecting the physical characteristics of each galaxy.  
The results from LVG modeling are qualitatively 
consistent with the theoretical predictions by \citet{krumholz07} and 
\citet{narayanan07b}.

Finally, we note that because the excitation of the 
CO~(3--2) line could be sensitive to temperature, 
less temperature dependent higher density 
gas tracers such as HCN~(1--0) or HCO$^{+}$ 
\citep[][but see \citet{papadopoulos07} for a counter argument for the use 
of HCO$^{+}$]{krumholz07,carpio08} 
could be the better tracer of star formation 
in quiescent galaxies, although the gas must be dense 
($n_{\rm H_2} \sim 10^{5-6}$~cm$^{-3}$) enough to excite the HCN emission.
These lines are also much weaker than CO lines and therefore harder 
to observe (see e.g. Figure~\ref{fig3}).  
While HCN emission is often used a tracer of dense gas 
feeding star formation activity, HCN emission from X-ray irradiated 
circumnuclear tori may 
become significant and therefore the HCN emission may be more pronounced 
towards AGN than starbursts in some cases \citep{kohno01}.  

\section{The evolutionary status of submillimeter galaxies}

\subsection{CO~(3--2) Source Size}

\begin{figure}
  \plotone{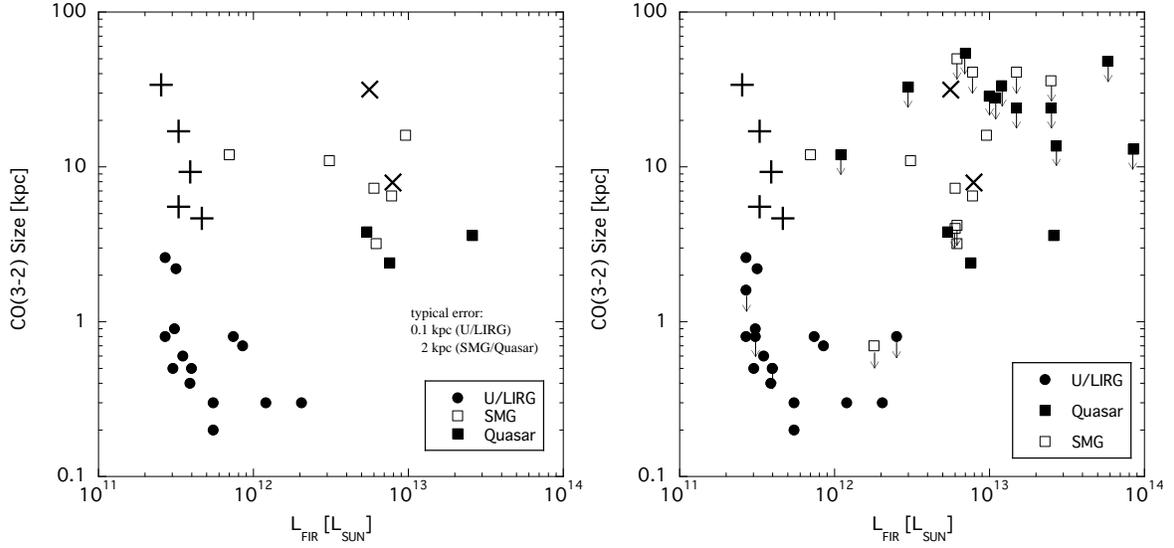}
  \caption{
($left$)  The CO~(3--2) size plotted against the FIR luminosities. 
    The CO~(3--2) minor axis was used for all sources. 
    We note that some of the LIRGs (especially the low luminosity ones) 
    are widely separated merging pairs, and here we have used the source sizes
    of each CO~(3--2) component.  Typical error bars are $\sim 0.1$~kpc for 
    the U/LIRGs and $\sim 2$~kpc for the high redshift sources.
We also plot the separation of the widely separated pairs for 
LIRGs (+) and SMGs (x).
($right$) Same as $left$ but with the galaxies with only upper limits to the
source size included for completeness.
  }
  \label{fig4}
\end{figure}

In \S 3, we have shown evidence that the overall  CO~(3--2) 
luminosity is well correlated with the total dusty star formation traced in 
FIR luminosities. 
In this section, we extend this argument and assume that the large scale  
spatial distribution of CO~(3--2) emission is also correlated with  
the extent of current star formation.  
To this end, we have plotted the CO~(3--2) emitting diameter (in kpc) 
against the FIR luminosities in Figure~\ref{fig4}. 
For the U/LIRGs, a trend is seen 
where the CO~(3--2) size decreases as a function of FIR luminosity.  
The CO~(3--2) size, however, can vary by an order of magnitude 
(0.3--3.1~kpc) for the LIRG population alone. 
In contrast, the ULIRGs are predominantly compact (0.4--1~kpc), 
and the large FIR luminosity seen in ULIRGs is likely related to this 
high central concentration of high density gas.   

This trend immediately breaks down in 
SMGs where the derived CO~(3--2) sizes (3--16~kpc) 
are roughly an order of magnitude 
larger than the ULIRGs. The average CO size for the SMGs ($8 \pm 2$
kpc) is only a factor of two smaller than
the average nuclear separation ($15 \pm 6$ kpc)of the pre-coalescence LIRGs
(e.g. VV~114, Arp~299; see also Figure~\ref{fig4}). 
%
The three resolved quasars have 
source sizes (2--4~kpc) that are systematically
smaller than the SMGs, but the sample size is too small for this trend 
to be conclusive.
We caution that, 
although the SMGs/quasars are observed with the highest angular 
resolution achievable with existing interferometers, these sources 
are only marginally resolved, and the resultant source sizes should be treated 
as a conservative upper limit to any internal structure present in these  
galaxies.  
The upper limits for the unresolved SMGs and quasars 
(see Figure~\ref{fig4}~($right$)) do not place very
useful limits on their intrinsic sizes, but are generally consistent
with the conclusions drawn from the resolved sources.

These sizes show that the star forming 
regions in local ULIRGs are much more compact than those in SMGs.  
This result is an interesting contrast to the analysis by \citet{bouche07} 
where they find that the millimeter or CO size 
of the SMGs themselves are much more compact than the optical 
size of the UV or optically selected population at $z\sim2$.  
The widely extended (as opposed to nuclear) star formation 
in SMGs was suggested previously using 
high resolution radio observations of spectroscopically identified SMGs 
by \citet{chapman04} who find 70\% of sources with size scale of 
$7 \pm 1$~kpc.  More recently, \citet{biggs07} 
combined the radio emission data of SMGs 
obtained using Multi-Element Radio Linked 
Interferometer Network (MERLIN) and VLA in the $u$-$v$ plane, and 
found an average SMG source size of 5~kpc, which is in agreement
with the results of \citet{chapman04} and also the
CO size measurements by \citet{tacconi06}.  
Two-arcsecond resolution SMA 
$890~\micron$ measurements \citep{iono06b,younger07,younger08a} 
of bright SMGs suggest source sizes of $< 8$~kpc, and 
new higher angular resolution SMA measurements 
suggest a smaller source size of $\sim 5$~kpc \citep{younger08b}. 
Thus, radio, CO, and submm observations all suggest that 
the size scale of the brightest SMGs are of order $\sim 5$~kpc, and this
is a factor of 2 -- 10  times larger than the CO~(3--2) 
derived source sizes in U/LIRGs.  
The average size of the SMGs in our CO (3--2) selected sample ($8 \pm
2$ kpc) is consistent with these numbers.

\subsection{Merger Scenarios for SMGs and quasars}

\begin{figure}
  \plotone{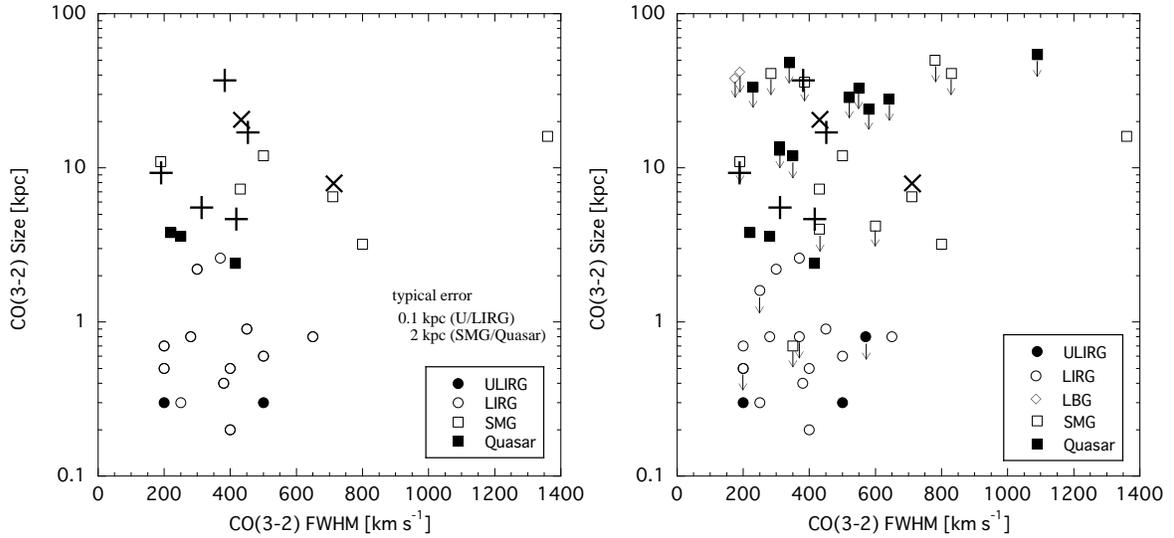}
  \caption{($left$) The relation between 
    source diameter (for SMGs, quasars, and LBGs) and the FWHM.  
($right$) Same as $left$ but with the galaxies with only upper limits to the
source size included for completeness.
  }
  \label{fig5}
\end{figure}

In Figure~\ref{fig5}, we plot the CO~(3--2) diameters as a function of
the CO~(3--2) FWHM.
Roughly two thirds (18/26) of the source sizes for SMG/quasars 
are only upper limits constrained from the observations. 
It is evident from Figure~\ref{fig5} that even though the size scales of SMGs
are comparable  to the nuclear separation of the widely separated LIRGs 
(i.e. + sign in Figure~\ref{fig5}), 
the FWHMs are on average 50\% larger (and a factor of two larger than
those of the ULIRGs).

The fact that both the average CO~(3--2) sizes and the SFEs are
similar for the SMGs and the local LIRGs suggests that they may be in
a similar evolutionary state, e.g., that many of the SMGs are in fact
early or intermediate rather than late stage mergers. Both the local
ULIRGs (which are known to be advanced mergers) and the quasars are
more compact (significantly so for the ULIRGs) and have SFEs that are
a factor of three larger than those of the SMGs.
In their analysis of several SMGs observed with the best angular
resolution, \citet{tacconi08} also conclude that SMGs are major
mergers in various evolutionary stages. They discuss three specific
SMGs (two of which, SMM~J163650+4057 and SMM~J123707+6214, are included
in our analysis) that each show two components separated by 8-25 kpc,
which is more characteristic of earlier stage mergers of which local
examples are Arp~299 and VV~114.
Given that most SMGs are not observed at quite such high resolution, 
it remains possible that many
SMGs could contain relatively compact CO reservoirs in two individual merging
components, but not be sufficiently separated to be resolved with the
existing data. However, this possibility would still be consistent
with our interpretation of the SMGs as predominantly earlier stage
mergers, as the resulting projected 
nuclear separation of SMGs 
would likely be as large as the widely separated LIRGs in our local sample. 

In mergers of two comparable mass galaxies, 
the FWHM derived from the total CO~(3--2) spectrum 
contains information pertaining to both galaxy rotation and encounter 
velocity.  In principle, the encounter velocity   
of two comparable mass \textit{field} galaxies in parabolic orbit 
is $\Delta V \sim \sqrt{2}~V_c$ where $\Delta V$ and $V_c$ are the
encounter velocity and circular velocity, respectively.
For local systems, $\Delta V$ amounts to $\sim 300$~km~s$^{-1}$ at maximum.
In reality, the observed $\Delta V$ is significantly smaller 
than this value, owing to, for example, projection and degeneracy in the 
merger phase, such as those seen in NGC~5331 ($\Delta V = 60$~km~s$^{-1}$) 
or NGC~5257/8 ($\Delta V = 11$~km~s$^{-1}$).  
At first glance, the broad CO~(3--2) linewidths of SMGs 
(i.e. 600~km~s$^{-1}$) argue against widely separated mergers.  
However, a simple analysis that assumes two equal mass field galaxies is 
likely inapplicable here because the brightest submillimeter sources 
are evidently embedded deep in the proto-cluster 
potential \citep{stevens03,greve05}. Thus, the encounter velocity of these 
galaxies could be significantly larger than those of local field galaxies.  
For comparison, a similarly large velocity dispersion 
is observed in the galaxy members of the Virgo Cluster \citep{rubin99}.  
The CO linewidth can trace the underlying stellar as well as the  
non-baryonic halo potential \citep{narayanan08},
and the suggested average stellar mass of SMGs 
\citep[$2.5 \times 10^{11}$~M$_{\odot}$;][]{borys05} is significantly 
more massive than a typical field galaxy in the local universe 
(i.e. the Milky Way). 
Therefore, we suggest that the larger CO~(3--2) 
FWHMs seen in SMGs  are likely due to a combination of an underlying
more massive dark matter potential and intrinsically more massive and
gas-rich merger progenitors.

Analysis of numerical simulations 
\citep{greve08,narayanan06,narayanan07a,narayanan08} have found that
the CO profile can display a variety of shapes owing to strong central
outflows which could appear as secondary emission peaks in the CO spectra.  
These simulations further
predict that the spectral characteristics have strong dependences 
on the sightline (i.e. the inclination of the source), as well as the
mass of the underlying potential especially toward the latter stage of the
evolution when the gas has settled in a disk (i.e. quasar phase).  
In addition, the simulated CO profile  evolves as a function of merger age, 
where kinematic asymmetries are seen primarily in the pre-coalescence
phase when the two galaxies  have not yet merged. In contrast, 
the spectrum becomes closer to Gaussian as the molecular gas becomes 
virialized and a quiescent molecular disk is formed.

While some of their adopted assumptions may not be best suited for 
analyzing the observational data presented in this study, 
these simulation results provide important physical insight.  
The narrower CO~(3--2) linewidths, and the  
dominance of Gaussian shapes in the quasar line profiles 
are consistent with the characteristics of a post-coalescence galaxy
seen in simulations.  
In our limited sample of only 
three quasars, the size scale of quasars with robust size estimates 
are intermediate between the nuclear separation of pre-coalescence LIRGs 
and the very compact ULIRGs
(see Figure~\ref{fig5}).
Given the factor of 10 larger average L$_{\rm FIR}$ in quasars
compared to the local ULIRGs, we might naively expect the starburst
region to be larger in the quasars, which could explain the size
difference to within a factor of  two.
Alternatively, the quasar activity can begin in one of the 
progenitor galaxies earlier in the 
merger evolution than that predicted in simulations, and the 
abundance of CO~(3--2) emitting molecular gas suggests intense concurrent 
starburst activity.  
It is also possible that CO~(3--2) emission entrained in powerful 
AGN jets can lead to an apparent increase in the CO~(3--2) size of quasars
\citep{narayanan06}. 
Further, the observed narrower CO linewidths of quasars may simply be due 
to an intrinsically less massive quasar disk \citep{greve05} or   
due to geometrical effects \citep{carilli06,wu07} with quasars
preferentially discovered in more face-on galaxies.
It is, however, impossible to obtain a robust conclusion 
from our current compilation of data which is limited in sample size, 
sensitivity, and angular resolution.  

Results from numerical simulations that include negative  
feedback from starbursts as well as the central AGN \citep[][]{hopkins05}    
suggest that SMGs and quasars are not in the same evolutionary stage, 
but linked through dynamical evolution where the quasar activity becomes  
visible during the final stages of the coalescence.  
We have provided evidence in this paper that the CO~(3--2) 
properties of many of the SMGs are consistent with the widely separated 
LIRGs in the local universe.   
A possible scenario suggested from these results is 
that SMGs are widely separated gas-rich mergers taking place 
near the centers of massive halos.  
Global disk-wide starburst activity in gas-rich high redshift mergers 
is also predicted theoretically by \citet{mihos01}.
On the other hand, some of the CO~(3--2) derived properties in
quasars, such as size and  SFE,
are more consistent with ULIRGs than with LIRGs or SMGs. 
Overall, the data summarized here are broadly consistent with this
evolution scenario.

Finally, we note that there is a selection effect in play.  There are 
interacting/merging galaxies in the local universe that are not U/LIRGs;
thus, the triggering of the U/LIRG activity also depends on the amount of 
pre-existing gas available in the merging pair.   
We also note that there are ULIRGs that are in the early stage of the 
interaction \citep{trung01}, as well as LIRGs that 
are apparently isolated (i.e. NGC~1068).  
There is also a case like the $z=4.7$ quasar -- SMG pair  BR1202--0725, in which
two submm bright sources \citep[one of which is a quasar, while the companion 
appears to harbor an obscured AGN,][]{iono06c} are separated by a projected 
distance of 26~kpc. The true evolution, therefore, is much more 
complicated than we envision in these simple scenarios, and 
sensitive high resolution ALMA surveys 
of SMGs, quasars, as well as quiescent high redshift galaxies in 
a variety of evolutionary stages should reveal their true nature.
Lastly, the gas consumption time for SMGs using the 
currently available fuel 
is estimated to be $\sim 40$~Myr \citep{greve05}, 
and this is significantly smaller than 
the typical major merger timescale of 1~Gyr.  If the SMGs and quasars 
indeed represent some stage of the evolution of a merger with sustained 
starbursts, then the dense 
gas must be replenished continuously in order to maintain 
a continuous burst of star formation throughout the merger evolution.  

\section{Summary}

In this paper, we have presented a detailed comparison of the CO~(3--2) 
emitting molecular gas 
between a local sample of luminous infrared galaxies (U/LIRGs) and a high 
redshift sample that comprises submm selected galaxies (SMGs), quasars,  
and two LBGs. The data for the local sample come from our
recent Submillimeter Array survey of CO~(3--2) emission in U/LIRGs while 
the CO~(3--2) data for the high redshift population are obtained from the
literature.  

(1) We find that 
L$^{'}_{\rm CO(3-2)}$ and L$_{\rm FIR}$ are
correlated over five orders of magnitude, which suggests that the
molecular gas traced in CO~(3--2) emission is a robust tracer of 
star formation activity.  The slope derived from the 
log~L$^{'}_{\rm CO(3-2)}$ -- log~L$_{\rm FIR}$ relation is 
$0.93 \pm 0.03$, and this near unity slope suggests that the 
efficiency of converting CO~(3--2) emitting molecular gas to  
massive stars is, to within a factor of two, 
nearly uniform across different types
of galaxies residing in different epochs. 
Within the local sample, 
the global star formation efficiency is lower in 
early to intermediate stage mergers than in the centers of more advanced 
stage mergers.  SMGs show lower star formation efficiencies that are
comparable to those of the widely separated LIRGs.

(2) The slope derived for the 
L$^{'}_{\rm CO(3-2)}$ -- L$_{\rm FIR}$ relation is 
significantly steeper than the slope derived from the 
same relation using the CO~(1--0) emission. We show that this 
slope approaches unity for higher critical density tracers.  
Further, we show that 
the CO~(3--2) emission in the central regions of U/LIRGs is, 
on average, nearly thermalized, whereas  the outskirts are 
subthermally populated.  

(3) From non-LTE LVG models as 
well as published merger evolution models that include radiative transfer, 
we argue that the CO~(3--2) line is a fairly good tracer of 
star formation in SMGs, where the star forming
gas density is  high enough that the dependence on temperature is 
relatively low.  
In contrast, L$^{'}_{\rm CO(3-2)}$ can show significant scatter
when the average gas density of the medium is lower than the critical density, 
which appears to occur more often in widely separated mergers in our sample. 
The scatter in the L$^{'}_{\rm CO(3-2)}$ -- L$_{\rm FIR}$ relation, 
therefore, reflects the characteristics of 
the galaxy (temperature, density, and AGN heating of dust in some
cases), but over a large range of  
luminosities, the CO~(3--2) line appears to be a good probe of star formation.

(4) The CO~(3--2) derived source sizes in U/LIRGs
show an apparent trend that the brighter ULIRGs are systematically 
more compact than the less FIR bright LIRGs.
The CO~(3--2) sizes of the SMGs are on average an order of magnitude 
larger than the ULIRGs, and are comparable to the separation of the
widely separated LIRGs in our sample. 
The three quasars with robust size constraints 
have source sizes that are somewhat smaller than
the SMGs 
and a factor of two larger than the ULIRGs.


(5) The similarity in the CO~(3--2) size and star formation efficiency
between SMGs and LIRGs suggests
that many of the SMGs studied here could be 
intermediate stage mergers.  
The SMG linewidths are, on average, much broader than many of the U/LIRGs, 
and we argue that the large encounter velocity likely arises from 
the massive halo potential in a proto-cluster environment. 
In contrast, quasars have 
smaller CO~(3--2) sizes and 
linewidths, and a dominance of Gaussian shapes in the line profiles.  These
characteristics are consistent with those seen in post-coalescence
galaxies in encounter simulations.


\acknowledgments

We thank the referee for detailed comments that improved the focus of
this paper.
DI would like to thank Ryohei Kawabe, Kotaro Kohno, 
Koichiro Nakanishi, Yoichi Tamura,  Akira Endo, Fumi Egusa, 
Bunyo Hatsukade, Masahiro Sameshima, 
Kazuyuki Muraoka, Shinya Komugi, Desika Narayanan for useful discussion, 
and Thomas Greve for kindly providing tabulated data for the SMGs.   
The Submillimeter Array is a joint project between the Smithsonian 
Astrophysical Observatory and the Academia Sinica Institute of Astronomy 
and Astrophysics and is funded by the Smithsonian Institution and the 
Academia Sinica. This research has made use of the NASA/IPAC Extragalactic 
Database (NED) which is operated by the Jet Propulsion Laboratory, 
California Institute of Technology, under contract with the National 
Aeronautics and Space Administration. 
This study was financially supported by MEXT Grant-in-Aid 
for Scientific Research on Priority Areas No. 15071202.
C.D.W.  acknowledges support by 
the Natural Science and Engineering Research Council of Canada (NSERC). 
A.J.B. acknowledges support by National Science Foundation grant AST-0708653.


%

%

\begin{deluxetable}{llccccccc}
\tabletypesize{\scriptsize}
\tablecaption{Sample Properties\label{table1}}
\tablehead{
\colhead{Source} & \colhead{type} & 
\colhead{Distance\tablenotemark{a}} &
\colhead{Lensing} &
\colhead{log L$_{\rm FIR}$\tablenotemark{b}} & 
\colhead{log L$_{\rm 1.4}$\tablenotemark{c}} & 
\colhead{S$_{\rm submm}$\tablenotemark{d}} & 
\colhead{Ref.\tablenotemark{e}} 
\\
& & [Mpc or $z$] & \colhead{Magnification} & [L$_{\odot}$] & 
[W~Hz$^{-1}$]& [mJy]
}
\startdata
\sidehead{\textbf{U/LIRGs}} 
IRAS 17208-0014 & ULIRG & 189        & \nodata & 12.41 & 23.55 & 48  & 1  \\
Mrk 231         & ULIRG & 179        & \nodata & 12.31 & 24.08 & 80  & 1  \\
Mrk 273         & ULIRG & 166        & \nodata & 12.08 & 23.68 & 56  & 1  \\
IRAS 10565+2448 & LIRG  & 191        & \nodata & 11.93 & 23.40 & 15  & 1  \\
UGC 5101        & LIRG  & 174        & \nodata & 11.87 & 23.79 & 37  & 1   \\
Arp 299         & LIRG  & 44 (4.7)   & \nodata & 11.74 & 23.20 & 101 & 1   \\
Arp 55          & LIRG  & 173 (9.2)  & \nodata & 11.60 & 23.12 & 26  & 1  \\
Arp 193         & LIRG  & 102        & \nodata & 11.59 & 23.11 & 39  & 1   \\
NGC 6240        & LIRG  & 107 (0.5)  & \nodata & 11.54 & 23.77 & 33  & 1  \\
VV 114          & LIRG  & 87 (5.6)   & \nodata & 11.50 & 23.36 & 26  & 1  \\
NGC 5331        & LIRG  & 145 (18.3) & \nodata & 11.49 & 22.99 & 27  & 1  \\
NGC 2623        & LIRG  & 80         & \nodata & 11.48 & 22.87 & 50  & 1  \\
NGC 5257/8      & LIRG  & 99 (38.4)  & \nodata & 11.43 & 23.03 & $<26$/104 & 1 \\ 
NGC 1614        & LIRG  & 69         & \nodata & 11.43 & 22.89 & 27  & 1 \\ 
\sidehead{\textbf{High Redshift Galaxies}} 
SMMJ02399-0136   & SMG/QSO & 2.808     & 2.5  & 12.98 & 24.57    & 9.6     & 2,3,4,5 \\ 
SMMJ0443007+0210 & SMG     & 2.509     & 4.4  & 12.25 & 23.10    & 1.6     & 4,5,6,7 \\
SMMJ123549+6215  & SMG     & 2.202     & ?    & 12.79 & 24.29    & 8.3     & 6,8 \\
SMMJ123707+6214  & SMG     & 2.49 (20) & ?    & 12.78 & 24.19    & 9.9     & 6,8 \\
SMMJ14011+0252   & SMG     & 2.565     & 5 (or 25-30) & 12.49 & 23.22 & 2.9& 9,10 \\
SMMJ16359+6612   & SMG     & 2.517     & 22   & 11.85 & \nodata  & 0.7     & 14,15  \\
SMMJ163650+4057  & SMG     & 2.385     & ?    & 12.89 & 24.83    & 8.2     & 6,16  \\
SMMJ163658+4105  & SMG     & 2.452     & 1    & 12.79 & 24.48    & 10.7    & 6,16,17,18 \\
SMMJ16371+4053   & SMG     & 2.380     & 1    & 12.89 & 24.36    & 10.5    & 17,19\\
SMMJ22174+0015   & SMG     & 3.099     & 1    & 12.79 & 24.39    & 6.3     & 17,20  \\
MIPSJ1428        & SMG     & 1.325     & ?    & 13.39 & 24.88    & 18.4    & 21,22\\
SMMJ163541+661144  & SMG     & 3.187   & 1.7    & 13.18 & \nodata  & 6.0    & 39 \\
SDSSJ1148+5251   & QSO     & 6.419     & ?    & 13.41 & 25.15    & 7.8     & 23,24\\
RXJ0911.4+0551   & QSO     & 2.796     & 22   & 12.06 & $<23.95$ & 1.2     & 25,26\\
LBQS1230+1627B   & QSO     & 2.735     & ?    & 13.39 & 24.94    & \nodata & 27\\
MG0414+0534      & QSO     & 2.639     & ?    & 13.19 & \nodata  & 16.7    & 26,28\\
LBQS0018         & QSO     & 2.620     & ?    & 13.09 & $<25.57$ & 17.2    & 29,30\\
VCVJ1409+5628    & QSO     & 2.583     & ?    & 13.93 & $<25.56$ & \nodata & 31\\
Cloverleaf       & QSO     & 2.558     & 11   & 12.88 & 24.36    & 5.4     & 26,32\\
IRAS F10214+4724 & QSO     & 2.286     & 17   & 12.73 & 23.23    & 2.9     & 33,34\\
SMMJ04135+1027   & QSO/SMG & 2.837     & 1.3  & 13.77 & $<25.72$ & 19.2    & 25,35 \\
HS1002+4400      & QSO     & 2.102     & ?    & 13.04 & \nodata  & 7.3\tablenotemark{f} & 36 \\
RXJ124913-055906 & QSO     & 2.247     & ?    & 12.85 & \nodata  & 7.2 & 36 \\
SMMJ131444+423814 & QSO/SMG& 2.556     & ?    & 12.48 & \nodata  & 3.0 & 36 \\
VV96 J140955+562827 & QSO  & 2.583     & ?    & 13.43 & \nodata  & 35.7\tablenotemark{f} & 36 \\
VV96 J154359+535903 & QSO  & 2.370     & ?    & 13.00 & \nodata  & 12.7\tablenotemark{f} & 36 \\
HS1611+4719 &         QSO  & 2.396     & ?    & 13.08 & \nodata  & 15.4\tablenotemark{f} & 36 \\
MS~1512-cB58     & LBG     & 2.727     & 31.8 & 10.99 & 22.85    & \nodata & 37  \\
J213512-010143   & LBG     & 3.074     & 8    & 11.48 & $<24.01$ & \nodata & 38  \\
\sidehead{\textbf{Average}\tablenotemark{g}} 
U/LIRGs          & & 129~Mpc   & \nodata & 11.74 & 23.37 & 49 \\
~~ULIRGs         & & 178~Mpc   & \nodata & 12.26 & 23.77 & 61 \\
~~LIRGs          & & 116~Mpc   & \nodata & 11.60 & 23.27 & 45 \\
SMGs             & & $z=2.493$ & 5.3     & 12.76 & 24.19 & 7.8 \\
Quasars          & & $z=2.782$ & 12.8    & 13.09 & 24.42 & 9.0 \\
LBGs             & & $z=2.898$ & 19.9    & 11.23 & 22.85 & \nodata \\
\enddata
\tablecomments{See Appendix of Paper~I for the discription of individual sources.  
}
\tablenotetext{a}{Values for U/LIRGs are given in D$_L$~(Mpc) 
  and the high redshift population  are given in redshifts.  For those 
  sources that are resolved into two or more galaxies, the projected
  nuclear separation derived from 2MASS K-band images are shown in () 
  in units of kpc.
}
\tablenotetext{b}{Values are
  corrected for lensing if the lensing factor is known.  See text for an explanation of how the 
  values are derived.  The FIR luminosity of cB58, and J213512 were 
  obtained from \citet{baker04} and \citet{coppin07}, respectively.}
\tablenotetext{c}{ 
  The 1.4~GHz data were obtained from the NVSS \citep{condon98} for the 
  U/LIRGs, \citet{smail02} \citet{chapman05} for the SMGs,  
  \citet{carilli04} \citet{white97} \citet{yun00}, \citet{barvainis96}, 
  for the quasars, and \citet{becker95} for the LBGs. Values are
  corrected for lensing if the lensing factor is known.
}
\tablenotetext{d}{SMA $880\micron$ flux for the U/LIRGs (paper~I).  SCUBA $850\micron$ flux 
  for the high redshift galaxies.  Values are corrected for lensing 
  magnification when the magnification factor is known.}
\tablenotetext{e}{1. Paper~I,   2. \citet{frayer98}, 3. \citet{genzel03}, 
4. \citet{smail97}, 5. \citet{smail02}, 6. \citet{tacconi06}, 
7. \citet{neri03}, 8. \citet{chapman05}, 9. \citet{frayer99}, 
10. \citet{smail98}, 11. \citet{swinbank04}, 12. \citet{motohara05}, 
13. \citet{smail05}, 14. \citet{sheth04},   15. \citet{kneib04}, 
16. \citet{scott02}, 17. \citet{greve05}, 18. \citet{ivison02}, 
19. \citet{greve04}, 20. \citet{barger99},  21. \citet{iono06a}, 
22. \citet{borys06}, 23. \citet{walter04}, 24. \citet{robson04}, 
25. \citet{hainline04}, 26. \citet{barvainis02}, 27. \citet{guilloteau99}, 
28. \citet{barvainis98},  29. \citet{svb05}, 30. \citet{priddey03},   
31. \citet{beelen04}, 32. \citet{weiss03}, 
33. \citet{downes95}, 34. \citet{rowan93},  35. \citet{knudsen03}, 
36. \citet{coppin08}, 37. \citet{baker04}, 38. \citet{coppin07}, 
39. \citet{knudsen08}
}
\tablenotetext{f}{ 
  Scaled the 1.2~mm flux in \citet{coppin08} to the $850\micron$ flux by assuming $\beta=1.5$.
}
\tablenotetext{g}{ 
  Excluding limits.
}
\end{deluxetable}

\begin{deluxetable}{lcccc}
\tabletypesize{\scriptsize}
\tablecaption{CO~(3--2) Derived Properties\label{table2}}
\tablehead{
  \colhead{Source}  & 
  \colhead{L$^{'}_{\rm CO(3-2)}$\dag} & 
  \colhead{$\Delta$ FWHM} &
  \colhead{Source Size\tablenotemark{a}\dag} & 
  \colhead{$(\frac{\rm L_{\rm FIR}}{\rm L^{'}_{\rm
        CO(3-2)}})$\tablenotemark{b}} 
}
\startdata
\sidehead{\textbf{U/LIRGs}} 
IRAS 17208-0014 & $4.5 \times 10^9$ & 570     & $<1.0 \times <0.8$ 
& 564~(2.75) \\ 
Mrk 231         & $2.6 \times 10^9$ & 200     & $0.5 \times 0.3$ 
& 793~(2.90) \\ 
Mrk 273         & $3.2 \times 10^9$ & 500     & $0.4 \times 0.3$ 
& 378~(2.58) \\ 
IRAS 10565+2448 & $1.9 \times 10^9$ & 200     & $1.0 \times 0.7$ 
& 439~(2.64) \\ 
UGC 5101        & $1.7 \times 10^9$ & 650     & $1.1 \times 0.8$ 
& 448~(2.65) \\ 
Arp 299 (total)\tablenotemark{c} & $1.3 \times 10^9$ & \nodata &
\nodata   
& 409~(2.61) \\ 
~~Arp 299E      & $8.4 \times 10^8$ & 400     & $0.3 \times 0.2$ 
& \nodata    \\ 
~~Arp 299W      & $3.1 \times 10^8$ & 250     & $0.4 \times 0.3$ 
& \nodata    \\ 
Arp 55 (total)\tablenotemark{c}  & $1.1 \times 10^9$ & \nodata &
\nodata    
& 349~(2.54) \\ 
~~Arp 55N       & $7.4 \times 10^8$ & 200     & $0.5 \times 0.5$ 
& \nodata    \\ 
~~Arp 55S       & $4.1 \times 10^8$ & 200     & $< 0.8 \times < 0.5$ 
& \nodata   \\ 
Arp 193         & $2.4 \times 10^9$ & 380     & $1.6 \times 0.4$ 
& 159~(2.20) \\ 
NGC 6240        & $7.4 \times 10^9$ & 500     & $0.9 \times 0.6$ 
& 47~(1.68)  \\ 
VV 114          & $3.1 \times 10^9$ & 300     & $3.0 \times 2.2$ 
& 102~(2.01) \\ 
NGC 5331 (total)\tablenotemark{c} & $2.6 \times 10^9$ & \nodata &
\nodata  
& 117~(2.07) \\ 
~~NGC 5331S     & $2.2 \times 10^9$ & 450     & $1.5 \times 0.9$ 
& \nodata    \\ 
~~NGC 5331N     & $4.3 \times 10^8$ & 370     & $<1.9 \times <0.8$ 
& \nodata   \\ 
NGC 2623        & $1.0 \times 10^9$ & 400     & $0.5 \times 0.5$ 
& 291~(2.46) \\ 
NGC 5257/8 (total)\tablenotemark{c} & $3.0 \times 10^9$ & \nodata &
\nodata  
& 93~(1.97)  \\ 
~~NGC 5257      & $6.7 \times 10^8$ & 250     & $<3.1 \times
<1.6$\tablenotemark{d} 
&  \nodata  \\ 
~~NGC 5258      & $2.2 \times 10^9$ & 370\tablenotemark{e} & $3.1
\times 2.6$\tablenotemark{d} 
&  \nodata  \\ 
NGC 1614        & $8.6 \times 10^8$ & 280     & $1.0 \times 0.8$ 
& 314~(2.50) \\ 
\sidehead{\textbf{High Redshift Galaxies}} 
SMMJ02399-0136      &$4.7 \times 10^{10}$  & $1360 \pm 50$ & 16
& 203~(2.31) \\ 
SMMJ0443007+0210    & $1.0 \times 10^{10}$ & $350 \pm 40$  & $< 3.1
\times < 0.7$
& 171~(2.23) \\ 
SMMJ123549+6215     & $4.2 \times 10^{10}$ & $600 \pm 50$  &
$<$4.2 (2.5)\tablenotemark{f} 
& 149~(2.17) \\ 
SMMJ123707+6214 (total) & $2.9 \times 10^{10}$ & \nodata   & \nodata
& 206~(2.31) \\ 
~~SMMJ123707+6214SW & $1.9 \times 10^{10}$ & $430 \pm 60$  & $7.3$
& \nodata    \\ 
~~SMMJ123707+6214NE & $1.0 \times 10^{10}$ & $430 \pm 60$  & $<4.0$
& \nodata    \\ 
SMMJ14011+0252      & $1.9 \times 10^{10}$ & $190 \pm 11$  & 11
& 164~(2.21) \\ 
SMMJ16359+6612      & $3.6 \times 10^{9}$\tablenotemark{g} & $500 \pm
100$          & $12 \times < 6$ 
& 196~(2.29) \\ 
SMMJ163650+4057     & $6.9 \times 10^{10}$ & $710 \pm 50$  & $6.5
\times <3.3$       
& 113~(2.05) \\ 
SMMJ163658+4105     & $5.6 \times 10^{10}$ & $800 \pm 50$  & $3.2$
& 110~(2.04) \\ 
SMMJ16371+4053      & $3.0 \times 10^{10}$ & $830 \pm 130$ & $<60
\times < 41$      
& 262~(2.42) \\ 
SMMJ22174+0015      & $3.7 \times 10^{10}$ & $780 \pm 100$ & $<69
\times <50$       
& 166~(2.22) \\ 
MIPS-J1428          & $1.4 \times 10^{11}$ & $386\pm 104$  & $< 36$
& 173~(2.24) \\ 
SMMJ163541+661144  & $2.8 \times 10^{10}$ & $284 \pm 50$ & $< 57 \times <41$ & 536 (2.73) \\
SDSSJ1148+5251      & $2.6 \times 10^{10}$ & $280$         & $3.6
\times <1.4$       
& 1000~(3.00) \\ 
RXJ0911.4+0551      & $5.1 \times 10^{9}$  & $350 \pm 60$  & $< 22
\times < 12$
& 222~(2.35) \\ 
LBQS1230+1627B      & $3.0 \times 10^{10}$ & \nodata       &
$<48 \times < 24$
& 816~(2.91) \\ 
MG0414+0534         & $9.2 \times 10^{10}$ & $580$         & $<24$
& 168~(2.23) \\ 
LBQS0018            & $5.4 \times 10^{10}$ & $163 \pm 29$  & \nodata
& 227~(2.36) \\ 
VCVJ1409+5628       & $7.9 \times 10^{10}$ & $311 \pm 28$  & $<20
\times <13$ 
& 1084~(3.03) \\ 
Cloverleaf          & $4.0 \times 10^{10}$ & $416 \pm 6$   & $<20
\times <12 (2.4)$
& 188~(2.27) \\ 
IRASF10214+4724     & $6.7 \times 10^{9}$  & $220 \pm 30$  & $ 3.8
\times < 2.5$
& 804~(2.91) \\ 
SMMJ04135+1027      & $1.7 \times 10^{11}$ & $340 \pm 120$ & $<96
\times < 48$
& 351~(2.55) \\ 
HS1002+4400         & $4.2 \times 10^{10}$ & $640 \pm 160$ & $<38
\times <28$       
& 262~(2.42) \\ 
RXJ124913-055906    & $3.6 \times 10^{10}$ & $1090 \pm340$ & $<73
\times <54$       
& 194~(2.29) \\ 
SMMJ131444+423814   & $1.2 \times 10^{10}$ & $550 \pm 220$ & $<42
\times <33$       
& 250~(2.40) \\ 
VV96 J140955+562827 & $8.2 \times 10^{10}$ & $310 \pm  30$ & $<19
\times <14$       
& 329~(2.52) \\ 
VV96 J154359+535903 & $3.1 \times 10^{10}$ & $520 \pm 140$ & $<46
\times <29$       
& 323~(2.51) \\ 
HS1611+4719         & $5.1 \times 10^{10}$ & $230 \pm  40$ & $<41
\times <33$       
& 235~(2.37) \\ 
MS1512-cB58         & $4.4 \times 10^{8}$  & $174 \pm 43$  & $<65
\times <38$~(2.0)\tablenotemark{h} 
& 224~(2.35) \\ 
J213512-010143      & $2.9 \times 10^{9}$  & $190 \pm 24$  & $<48
\times <42$       
& 105~(2.02) \\ 
\sidehead{\textbf{Average}\tablenotemark{i}} 
U/LIRGs  & $(2.6 \pm 0.5) \times 10^{9} $ & $360\pm30$  &
$(1.1\times0.8)\pm0.3$ 
& $322 \pm 56$ \\ 
~~ULIRGs & $(3.4 \pm 0.6) \times 10^{9} $ & $420\pm110$ &
$(0.5\times0.3)\pm0.1$ 
&$578 \pm 120$ \\ 
~~LIRGs  & $(2.4 \pm 0.6) \times 10^{9} $ & $350\pm30$  &
$(1.3 \times 0.9) \pm0.3$ 
& $252 \pm 46$ \\ 
SMGs     & $(4.4 \pm 1.1) \times 10^{10}$ & $590\pm90$  & $8.4 \pm
1.9$          
& $194 \pm 20$ \\ 
Quasars  & $(5.0 \pm 1.0) \times 10^{10}$ & $430\pm60$  & $3.3 \pm
0.4$          
& $430 \pm 83$ \\ 
LBGs     & $(1.6 \pm 1.2) \times 10^9   $ & $180\pm10$  & \nodata        
& $170 \pm 60$ \\ 
\enddata
\tablecomments{
  Units are 
  [K~km~s$^{-1}$~pc$^2$] for log L$^{'}_{\rm CO(3-2)}$, 
  [km~s$^{-1}$] for $\Delta$~FWHM, 
  [kpc] for source size, and
  [L$_{\odot}$~(K~km~s$^{-1}$~pc$^{-2}$)$^{-1}$)] for L$_{\rm
    FIR}$/L$^{'}_{\rm CO(3-2)}$ in linear and logarithmic scales.
}
\tablenotetext{a}{
  The deconvolved source size.  Gaussian fits for IRAS~17208-0014,
  Arp~55S, NGC~5331N, NGC~5257 give only the upper limits shown.
  Sizes derived from CO transitions other 
  than CO~(3--2) are used for some of the high
  redshift sources.  For high redshift sources with known luminosity  
  magnification factors, we have scaled the CO size by the square root 
  of the luminosity magnification assuming that the surface brightness is
  conserved during gravitational lensing.  
}
\tablenotetext{b}{
  The logarithm of the values are shown inside ().
}
\tablenotetext{c}{
  For the sources with a ``total'' entry, we divide the sum of the gas masses of the 
  two components by the total FIR luminosity to obtain 
L$_{\rm    FIR}$/L$^{'}_{\rm CO(3-2)}$.
}
\tablenotetext{d}{
  The CO~(3--2) source sizes in NGC~5257/8  mainly arise 
  from the bright arms rather than a central concentration. 
}
\tablenotetext{e}{
  The total spectrum of NGC~5258 has two components, and we have listed here the
  approximate FWHM of the entire spectrum.  
}
\tablenotetext{f}{
  The CO~(6--5) source size.
}
\tablenotetext{g}{
  Only the brightest component detected in CO~(3--2) by \citet{sheth04}.
}
\tablenotetext{h}{
  The size of UV emitting region \citep{baker04}.
}
\tablenotetext{i}{
Mean $\pm$ uncertainty in the mean, excluding limits. 
The standard deviation is a factor
of $\sqrt{N-1}$ larger than the uncertainty in the mean, where $N$ is
the number of galaxies in the subsample.  
}
\tablenotetext{\dag}{
  Quantities that depend on gravitational lensing.
}
\end{deluxetable}

\begin{deluxetable*}{lcccccccccc}
\tabletypesize{\scriptsize}
\tablewidth{0pt}
\tablecaption{CO~(3--2) -- CO~(1--0) Ratios in U/LIRGs\label{table3}}
\tablehead{
\colhead{Source}  & 
\colhead{$R_{total}$\tablenotemark{a}} & 
\colhead{$R_{peak}$\tablenotemark{b}} &
\colhead{CO~(1--0) Refs.\tablenotemark{c}} 
}
\startdata
IRAS 17208-0014 & 0.40 & 0.41 & 1 \\
Mrk 231 & 0.35 & 0.82 & 1 \\
Mrk 273 & 0.63 & \nodata\tablenotemark{d} & 1 \\
IRAS 10565+2448 & 0.33 & 0.72 & 1 \\
UGC 5101& 0.46 & 0.48  & 2 \\
Arp 299E& 0.74 & 1.11 & 3\\
Arp 299W& 1.25 & 1.30 & 3\\
Arp 55N& \multirow{2}{*}{0.20}& \nodata\tablenotemark{e} & 4\\
Arp 55S&  & \nodata\tablenotemark{e} & 4\\
Arp 193& 0.61& $<1.26$ & 1\\
NGC 6240 & 0.83 & 3.21 & 5 \\
VV 114& 0.25& 0.49 & 6 \\
NGC 5331S & 0.30 & 0.55 & 7\\
NGC 5331N & 0.50 & 0.50 & 7 \\
NGC 2623& 0.44 & 0.74 & 5\\
NGC 5257 & 0.21 & 0.36 & 7\\
NGC 5258 & 0.38 & 1.06 & 7\\
NGC 1614& 0.35 & \nodata\tablenotemark{f} & 8\\
average & 0.48  & 0.93 \\
\enddata
\tablenotetext{a}{
  The CO~(3--2) to CO~(1--0) flux ratio using the total luminosities.
}
\tablenotetext{b}{
  The CO~(3--2) to CO~(1--0) flux ratios using the peak fluxes.
  The peak fluxes were derived after convolving the CO~(3--2) maps 
  to the lower resolution CO~(1--0) maps, 
  except for Arp~193 where the CO~(1--0) map has higher angular resolution
  than the CO~(3--2) map.  The peak CO~(1--0) fluxes were obtained from the
  images published in the references.  These ratios trace the inner
  1--3~kpc of the galaxies.
}
\tablenotetext{c}{
  1. \citet{downessolomon98}, 2. \citet{genzel98}, 3. \citet{aalto97}
  4. \citet{sanders88b}, 5. \citet{bryant99}, 6. \citet{yun94}
  7. \citet{iono05}, 8. \citet{scoville89}
}
\tablenotetext{d}{
  There appear to be inconsistencies in the published values   
  of Mrk~273 in Fig.~15 of \citet{downessolomon98}.  
  The peak flux level ($\sim144$~Jy~km~s$^{-1}$~beam$^{-1}$) 
  exceeds the total flux ($78$~Jy~km~s$^{-1}$) 
  given in the same paper.
}
\tablenotetext{e}{
  Although the CO~(1--0) contours are shown overlaid on an 
  optical image, the contour levels were not explicitly 
  given in \citet{sanders88b}.
}
\tablenotetext{f}{
  NGC~1614 is unresolved in this paper.
}
\end{deluxetable*}

\end{document}